\documentstyle[preprint,aps]{revtex}
\begin{document}
\draft
\def\a{\alpha}
\def\b{\beta}
\def\s{\sigma}
\def\t{\tau}
\def\d{\delta}
\def\be{\begin{equation}}
\def\ee{\end{equation}}
\def\bea{\begin{aqnarray}}
\def\eea{\end{eqnarray}}

\title{Duality Relations for Potts  Correlation Functions}
\author{F. Y. Wu}
\address{Department of Physics,
	 Northeastern University, Boston, Massachusetts 02115, U. S. A. \\
Centre Emile Borel - UMS 839 IHP (CNRS/UPMC) - Paris, France}
\maketitle

\begin{abstract}
Duality relations are obtained 
for  correlation functions  of the $q$-state
Potts model on any planar lattice or graph using a simple
graphical analysis.  For the two-point correlation  we show that
the correlation length is precisely the  surface tension of the dual
model, generalizing a  result known to hold   for the Ising model.  For 
the three-point
correlation an explicit expression is obtained
relating the
correlation function to ratios of  dual partition functions
under fixed boundary conditions.

\end{abstract}
\bigskip
\bigskip

\newpage   
\section{Introduction}
It is now well-known   \cite{watson,zia} that the 
correlation length  
of the Ising model in two dimensions is 
precisely the surface tension of the dual lattice.   
It is also known by folklore (see, for example, \cite{wu})
that a similar duality exists for the Potts model \cite{potts}.  
However, 
detailed discussion of the correlation 
duality for the Potts model has yet to appear 
in the literature.  In view of the important role played by such
duality relations 
in discussions of the equilibrium crystal
shapes \cite{wortis,shapes}, 
a definitive understanding of
this subject matter is clearly needed.

Here we take up this  question
and consider more generally the 
duality for the $n$-point correlation function.
On the basis of a simple graphical analysis, we 
derive 
duality relations for the two- and three-point correlation 
functions. 
 Particularly, we establish that the correlation length 
in the large lattice limit is precisely
 the surface tension of the dual model,
thus generalizing the known Ising result, and that 
the three-point correlation function is
given by similarly defined expressions
in the dual space.
Our analysis, which does 
not concern with the internal structure of the underlying
lattice, is quite general and applies to the Potts model
with  arbitrary edge-dependent interactions  and/or on 
any planar lattice or graph.
It can also be extended to higher correlations.
 
Consider a $q$-state Potts model \cite{potts} 
on a two-dimensional  lattice  $L$
with  a free boundary. 
 An example of $L$ is 
the rectangular $4\times 6$ lattice
shown in Fig. 1(a), but 
  more generally 
$L$ can be
any planar lattice or graph
which does not need to be regular.  The site $i$
of $L$ is occupied by  a  spin $\s_i$ 
which can take on values
$1,2,...,q$.
Two spins at sites $\alpha$ and $\beta$ 
and in states $\s_\alpha$ and $\s_\beta$ interact with  
a Boltzmann factor ${\rm exp}
[K_{\alpha\beta}\d(\s_\alpha,\s_\beta)]$. 
Let  $i, j, \cdots, n$ denote {\it any} $n$ sites on the boundary. 
We introduce the $n$-point correlation function as
the probability that 
the spins at sites $\{i,j,\cdots, n\}$  
are in the respective states $\{\s, \s', \cdots, \s^{(n-1)}\}$, 
\be
P_n(\s, \s', \cdots, \s^{(n-1)} ) = <\d(\s_i, \s) \d(\s_j, \s')
\cdots \d(\s_{n}, \s^{(n-1)})> , \label{n}
\ee
where $<\cdot>$ denotes statistical averages. 
It is sometimes
convenient  to consider the correlation 
function (see, for example \cite{wuwang,wu} for $n=2$)
\begin{eqnarray}
\Gamma_n(\s_i, \s_j, \cdots, \s_{n} )& =& 
< q^{n-1}\d(\s_i, \s_j) \d(\s_j, \s_k)
\cdots \d(\s_{n-1}, \s_{n})-1>  \nonumber \\
&=& q^n P_n (\s, \s, \cdots, \s) -1
, \label{correlation}
\end{eqnarray}
a quantity 
which vanishes identically if the $n$ spins 
$\{\s_i,\s_j,\s_k \cdots, \s_{n-1},\s_n\}$ 
are completely uncorrelated.
In the case of $q=n=2$, for example, one writes 
$2\d(\s_i, \s_j) = 1+\s_i\s_j$ where $\s_{i,j} = \pm 1$,
then (\ref{correlation}) becomes  the 
usual expression $\Gamma_2(\s_i, \s_j) =<\s_i\s_j>$ for the Ising model.
It is clear that the correlation 
function $P_n$ 
is more general than 
$\Gamma_n$.

Our analyses is based on the repeated use of a
fundamental duality relation
for the Potts model  \cite{wuwang}.
Construct $L^*$, the dual lattice (graph) of $L$, 
by
placing spins in the {\it faces} of $L$ including the exterior one.
In Fig. 1(a), for example,
the faces of the dual lattice are denoted by crosses 
and the letter $s$, so $L^*$ has
$N^*= 15+1=16$ sites.  
Edges of $L^*$ (not shown in Fig. 1(a))
bisect, and are in one-one correspondence with, edges of $L$.
Let $K_{\alpha\beta}^*$ be the 
interaction  of the edge bisecting 
the interaction $K_{\alpha\beta}$.  Let $Z(\{K_{\alpha\beta}\})$
and $Z^*(\{K_{\alpha\beta}^*\})$ be the respective
partition functions on $L$ and $L^*$.
Then one has the following  fundamental duality
relation \cite{wuwang}
\be
Z(\{K_{\alpha\beta}\})
=qC Z^*(\{K^*_{\alpha\beta}\}) \label{duality}
\ee
where $K_{\alpha\beta}$ and $K^*_{\alpha\beta}$ are related by
\be
\bigl(e^{K_{\alpha\beta}}-1\bigr)
\bigl(e^{K^*_{\alpha\beta}}-1\bigr) =q, \label{dual}
\ee
and $C=q^{-N^*} \prod_{\rm edges} \bigl(
e^{K_{\alpha\beta}}-1\bigr)$.  

\section{The two-point correlation function}

Let $i$ and $j$ denote any two sites on the boundary of $L$
as shown in Fig. 1(a).
Further let
\be
Z_{11}= Z(\s_i=\s_j=1), \hskip 1cm Z_{12}=Z(\s_i=1, \s_j=2) 
\label{partialZ}
\ee
be the  partition functions with $\s_i$ and $\s_j$  fixed in definite states
$1$ and/or $2$,
Then, by symmetry we have
\be 
Z(\{K_{\alpha\beta}\})= q Z_{11} + q(q-1) Z_{12}. \label{partitionfunction}
\ee
Clearly, we also have $P_2(1,1)=Z_{11}/Z$ 
and $P_2(1,2)=Z_{12}/Z$.

 Let $Z^*_{11}$ denote the  dual partition function with
the spin in the exterior face in a  definite  state, say, $s$.
Alternately, from Fig. 1(a),  we see that $Z^*_{11}$   can 
also be regarded as
the dual partition function with all boundary spins interact
with the exterior spin fixed at state $s$. 
Then we have
\be
 Z^* (\{K_{\alpha\beta}\}) = qZ^*_{11}. \label{dualpart}
\ee
Substituting (\ref{partitionfunction}) and (\ref{dualpart}) into the duality
relation (\ref{duality}), one obtains the relation
 \be
Z_{11} +(q-1) Z_{12} = qC Z^*_{11}  . \label{equation1}
\ee

To obtain a second  relation relating $Z_{11}$ and $Z_{12}$, we 
 apply the duality relation (\ref{duality}) to 
 a lattice, or graph, constructed  from $L$ by adding one additional
edge connecting sites $i$ and $j$
with an interaction $K$, a situation
 shown in Fig. 1(c).
Clearly, the respective partition functions for the new
lattice and its dual are now
\be
\tilde Z= qe^{K} Z_{11} + q (q-1) Z_{12} \label{p1}
\ee
and  
\be 
\tilde Z^* = q e^{K^*} Z^*_{11} + q (q-1) Z^*_{12} \label{p2}
\ee
where $Z^*_{12}$ is the  partition function of the dual 
under the boundary condition such that  the spins in the  two
exterior faces are in definite states $s\not= s'$
as shown in Fig. 1(b).
Here, $K$ and $K^*$ are related by (\ref{dual}).
 Substituting (\ref{p1}) and (\ref{p2}) into (\ref{duality}) and noting that
$N^*$ is increased by $1$ in the new lattice,
 one obtains a second relation
\be
e^{K}Z_{11} +(q-1) Z_{12} = C(e^{K}-1) [e^{K^*}Z^*_{11} +(q-1)Z^*_{12}].
  \label{equation2}
\ee

We now solve (\ref{equation1}) and (\ref{equation2}) for $Z_{11}$ and $Z_{12}$.
This leads to, after using (\ref{dual}) to eliminate $e^{K^*}$,
\begin{eqnarray}
Z_{11} &= &C [Z^*_{11} +(q-1)Z^*_{12}] \nonumber \\
Z_{12} &= &C (Z^*_{11} -Z^*_{12}).\label{solution}
\end{eqnarray}
Note that the interaction $K$ introduced to
facilitate calculations does not enter (\ref{solution}). 
The expression (\ref{solution}) now leads to  the
desired expression for the correlation function.
  Particularly,
using the identity $Z=q^2CZ^*_{11}$, we obtain
 \be
P_2(\s, \s') = {1\over {q^2}} \biggl[
1+ (q\d_{\s,\s'} -1) {{Z^*_{12}}\over {Z^*_{11}}} \biggr]\label{result1}
\ee
and
\be
\Gamma_2(\s_i, \s_j) = 
(q-1) \biggl({{Z^*_{12}}\over {Z^*_{11}}}\biggr). \label{result}
\ee

For the Potts model the surface tension $\t$  is defined by \cite{fg} 
\be
\t = -\lim_{D\to \infty}{1\over D} \ln \biggl( {{Z^*_{12}}\over {Z^*_{11}}} \biggr)  \label{tension}
\ee
where $D$ is the distance between  sites $i$ and $j$, the two points where
the boundary spin state changes from $s$ to $s'$ in Fig. 1(b).  Our 
result (\ref{result}) now relates
the correlation function on $L$  to the surface tension on $L^*$. 
Particularly, the known exponential decay  $e^{-D/\xi}$ of the two-point correlation
$\Gamma_2$ above the transition temperature $T_c$ \cite{hintermann},
where $\xi$ is the correlation length, leads to the identity
$\xi^{-1} = \t$,
where $\t$ is  the  surface tension 
 of the dual model below $T_c$.
This 
 generalizes the corresponding
$q=2$ result of the Ising model \cite{watson,zia} to all $q$.

\section{The three-point correlation}

Consider any three sites $i,j,k$ on the boundary of $L$ as shown in
Fig. 2.  Here, to emphasize the generality of our consideration
and the fact that  the  internal structure of $L$ does not enter the picture,
the lattice is shown symbolically as a shaded region,
Let
\be
Z_{\s\s'\s''} = Z(\s_i=\s, \s_j=\s', \s_j=\s'') .
\hskip 1cm \s, \s', \s'' = 1,2,...,q \label{threespin}
\ee
denote the partition function with 
$\s_i,\s_j,\s_k$ fixed in definite states.
Then, in analogous to (\ref{partitionfunction}) the partition function can be written 
as
\be
Z= q Z_{111} + q(q-1)[ Z_{211} +Z_{121} +Z_{112}] + q(q-1)(q-2)Z_{123}. \label{partfunc1}
\ee
In a similar manner, let $Z^*_{ss's''}$ be the  partition function
of the dual model under
a boundary condition such that all boundary spins  between sites $i$ and
$j$ of $L$ interact with a fixed spin state $s''$, 
all boundary spins between $j$ and $k$
interact with a spin  state $s$, and all boundary spins between $k$ and $i$
interact with a spin  state $s'$, a situation shown in Fig. 2.
Then,  the dual partition function can be written as
\be
Z^* = q Z^*_{111}.  \label{dualpart1}
\ee
Substituting (\ref{partfunc1}) and (\ref{dualpart1}) into (\ref{duality}), one obtains
the relation
\be
Z_{111} + (q-1)( Z_{211} +Z_{121} +Z_{112}) + (q-1)(q-2)Z_{123}
= qC  Z^*_{111}.  \label{relation1}
\ee

Next we modify $L$ by connecting sites
$j$ and $k$ with a new edge and apply the 
result (\ref{solution}).
In the first line of (\ref{solution}), one has, by definition,
\begin{eqnarray}
Z_{11} &=& \sum_{i=1}^q Z_{i11} = Z_{111} +(q-1) Z_{211} \nonumber \\
Z^*_{11} &=& Z^*_{111} \nonumber \\
Z^*_{12} &=& Z^*_{122} = Z^*_{211}. \label{appl}
\end{eqnarray}
Substituting
(\ref{appl}) into (\ref{solution}), one obtains a second relation
\be
Z_{111} +(q-1) Z_{211} =C[Z^*_{111} +(q-1) Z^*_{211}]. \label{relation2}
\ee
 In a similar manner
by connecting sites $\{k,i\}$ or $\{i,j\}$ with a new edge, we obtain, respectively,
 \begin{eqnarray}
&&Z_{111} +(q-1) Z_{121} =C[Z^*_{111} +(q-1) Z^*_{121}]\label{relation3} \\
&&Z_{111} +(q-1) Z_{112} =C[Z^*_{111} +(q-1) Z^*_{112}] \label{relation4} 
 \end{eqnarray}
Finally, we apply the duality relation (\ref{duality}) to a  lattice
constructed 
by introducing 3 new edges connecting  sites $i,j$ or $k$ 
to a common new  point with interactions $K$ as
shown in Fig. 2. This process increases  $N^*$ by $2$ and yields
the duality relation
\begin{eqnarray}
&&(e^{3K}+q-1)Z_{111} +(q-1)(e^{2K}+e^K+q-2)(Z_{211}+Z_{121}+Z_{112}) 
\nonumber \\
&& \hskip 3cm +(q-1)(q-2)(3e^K+q-3)Z_{123} \nonumber \\
&&\hskip 1cm =C (e^K-1)^3 q^{-1}[e^{3K^*} Z^*_{111} + 
(q-1) e^{K^*} (Z^*_{211}+Z^*_{121}+Z^*_{112})\nonumber \\
 && \hskip 3cm +(q-1)(q-2) Z^*_{123}]. \label{relation5}
\end{eqnarray}

From (\ref{relation1}),  (\ref{relation2}),  (\ref{relation3}),  
(\ref{relation4}),  and (\ref{relation5}) and using
(\ref{dual}) to eliminate $K^*$, we  solve  
$Z_{111}$, $Z_{211}$, $Z_{121}$, $Z_{112}$, and $Z_{123}$ 
in terms of the dual partition functions $Z^*_{ss's''}$.
 This leads to 
 \begin{eqnarray}
Z_{111}&=&( C/q)[  Z^*_{111} +(q-1)(q-2)
  Z^*_{123}+(q-1)(Z^*_{211}+ Z^*_{121}+ Z^*_{112})] \nonumber \\
Z_{123}&= &(C/q)[  Z^*_{111} + 2
  Z^*_{123}- Z^*_{211}- Z^*_{121}- Z^*_{112}] \nonumber \\
Z_{211} &=&  
(C/q) [ Z^*_{111}-(q-2) Z^*_{123} +(q-1)Z^*_{211}-Z^*_{121}-Z^*_{112}] \nonumber \\
Z_{121} &=&  
(C/q) [ Z^*_{111}-(q-2) Z^*_{123} -Z^*_{211}+(q-1)Z^*_{121}-Z^*_{112}]\nonumber \\
Z_{112} &=&  
(C/q) [ Z^*_{111}-(q-2) Z^*_{123} -Z^*_{211}-Z^*_{121}+(q-1)Z^*_{112}] .
   \label{three}
\end{eqnarray}
This result is again independent of the parameter $K$ used in the evaluation.
 
Using (\ref{three})
and the identity
$Z=Cq Z^*=Cq^2 Z^*_{111}$, we finally  obtain
\begin{eqnarray}
&&P_3(\s,\s',\s'')
= {1\over {q^3}} \biggl[ 1 + 2 p_{123} -( p_{211} + p_{121}  +p_{112})
 +q ( p_{211} - p_{123}) \d_{\s', \s''} \nonumber \\
&&\hskip 1cm +
q ( p_{121} - p_{123}) \d_{\s'', \s}  
+q ( p_{112} - p_{123}) \d_{\s, \s'}
+ q^2 p_{123} \d_{\s,\s'}\d_{\s',\s''} \biggr]
\label{threec}
\end{eqnarray}
and
\begin{eqnarray}
\Gamma_3(\s_i, \s_j, \s_k)& =& (q-1)
\biggl[p_{112}+p_{121}+p_{211}+(q-2)p_{123}\biggr]
   \nonumber \\
&=&\Gamma_2(\s_i,\s_j) +\Gamma_2(\s_j,\s_k) +\Gamma_2(\s_k,\s_i) +
(q-1)(q-2)p_{123}.
\end{eqnarray}
 Here,  
\be
 p_{ss's''} = {{Z^*_{ss's''}}/{Z^*_{111}}}
\ee
is the ratio of dual partition functions which 
can  be interpreted as  appropriately 
defined
surface tensions, and  we have used $(q-1)p_{112} =
\Gamma_2(\s_i, \s_j)$, etc.  Since $Z^* = qZ^*_{111}$, 
the ratio $p_{ss's''}$ gives also the probability
that the particular boundary condition $\{s, s', s''\}$ of Fig. 2 occurs in
the dual partition function $Z^*$.  It is readily verified that
the identity  
\be 
\sum_{\s''}
P_3(\s,\s',\s'') = P_2(\s,\s')
\ee
is satisfied.

\section{Summary}
In summary, we have presented a graphical analysis of 
 duality relations leading to explicit expressions for  correlation functions
of the $q$-state Potts model on any planar lattice
or graph.  For the two-point correlation 
our result (\ref{result}) generalizes a relation previously known  for the
Ising model.  For three-point correlations, 
our result (\ref{threec}) is new and  can be used to 
compute  any three-point correlation. 
In all cases the correlation functions are found to be given by  
ratios of dual partition functions under  fixed boundary conditions, which
 can in turn be interpreted as appropriately defined
surface tensions.
Our consideration can be extended  in a straightforward fashion 
to    
 higher correlations $P_n$ and $\Gamma_n$ for $n>3$, 
 and to the more general  $(N_\alpha, N_\beta)$ model \cite{domany}
including the  $N_\alpha=N_\beta=2$ Ashkin-Teller model  \cite{at}.
  
\section{Acknowledgements}

I would like to thank R. K. P. Zia for suggesting this problem and 
enlightening discussions. I would also like to thank 
M.-J. Maillard for the hospitality at the
Institut Henri Poincar\'e and H. Kunz for the 
hospitality at Institut de Physique Th\'eorique, Lausanne, where
this research was initiated.  
This work is supported in part by the National Science Foundation
Grant DMR-9614170.

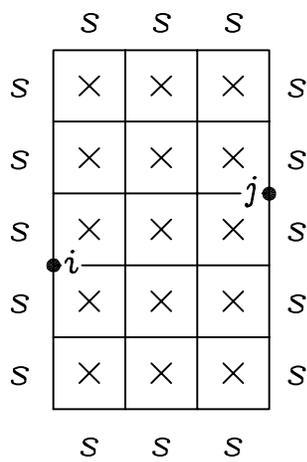
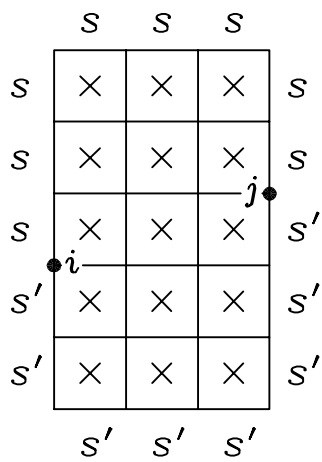
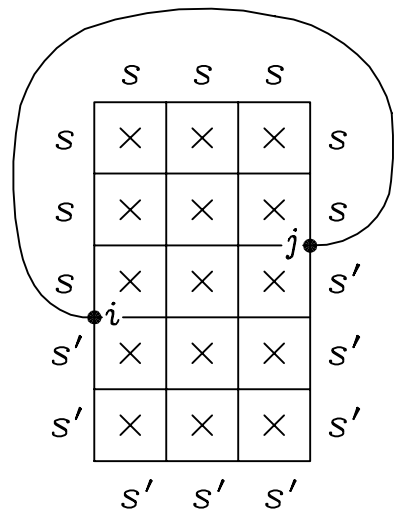
\begin{figure}
\caption{(a) A $4\times 6$ lattice $L$. 
The dual spins are denoted by crosses 
and the letter $s$. 
(b) The boundary condition ($s\not=s'$) for $Z^*_{12}$.
(c) The addition of one edge to $L$ connecting sites $i$ and $j$.}
\end{figure}

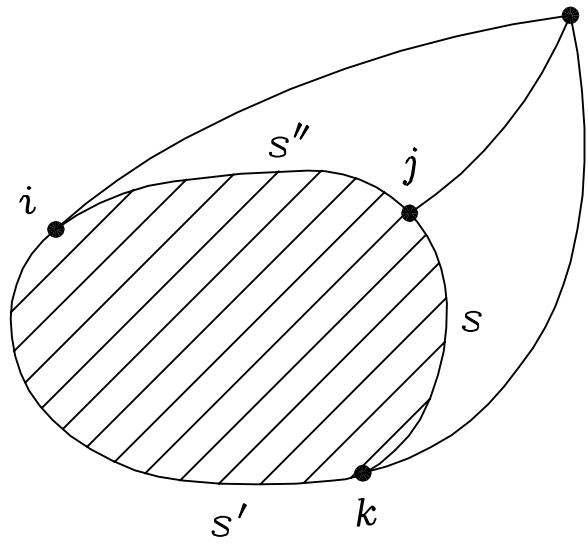
\begin{figure}
\caption{The addition of 3 edges to $L$ 
(the shaded region) connecting sites $i$, $j$, and $k$ to a common
point.}

\end{figure}

\end{document}